\documentclass[10pt,a4paper]{article}

\usepackage[utf8]{inputenc}
\usepackage[english]{babel}

\usepackage{bussproofs}
\usepackage{amsthm} 
\usepackage{amsfonts}
\usepackage {tikz}
\usepackage{mathrsfs}

\usepackage{amssymb}
\usepackage{graphicx}

\usetikzlibrary{fit,shapes.geometric, arrows}
\usetikzlibrary{positioning}
\usetikzlibrary{calc,shapes.multipart,chains}
\definecolor {processblue}{cmyk}{0.96,0,0,0}

\usepackage{subcaption}

\theoremstyle{definition} 
\newtheorem{definition}{Definition}[section] 
\theoremstyle{plain} 

\newtheorem{example}{Example}

\usepackage[scr=boondox,  
            cal=esstix]   
           {mathalpha}

\usepackage{appendix} 

\usepackage{quoting} 
\quotingsetup{font=small}

\usepackage[square,sort,comma]{natbib}
\usepackage{varioref} 

\usepackage{float}

\usepackage{cancel}

\usepackage{hyperref} 
\usepackage{geometry}
\usepackage{bussproofs}
\usepackage{amsthm} 
\usepackage{amsfonts}
\usepackage {tikz}

\usetikzlibrary{fit,shapes.geometric, arrows}
\usetikzlibrary{positioning}
\usetikzlibrary{calc,shapes.multipart,chains}
\usepackage {xcolor}
\definecolor {processblue}{cmyk}{0.96,0,0,0}

\usepackage{comment}

\usepackage{subcaption}

\newcommand*{\nc}[2]{#1\mathbin{\left| \sim \vphantom{#1#2} \right.}#2}%

\makeatletter

\newcommand{\tright}{\mathrel\triangleright}
\DeclareRobustCommand{\btleft}{\mathrel{\mathpalette\btlr@\blacktriangleleft}}
\DeclareRobustCommand{\btright}{\mathrel{\mathpalette\btlr@\blacktriangleright}}

\newcommand{\btlr@}[2]{%
  \begingroup
  \sbox\z@{$\m@th#1\triangleright$}%
  \sbox\tw@{\resizebox{1.1\wd\z@}{1.1\ht\z@}{\raisebox{\depth}{$\m@th#1\mkern-1mu#2$}}}%
  \ht\tw@=\ht\z@ \dp\tw@=\dp\z@ \wd\tw@=\wd\z@
  \copy\tw@
  \endgroup
}

\newcommand*{\Scale}[2][4]{\scalebox{#1}{$#2$}}%

\makeatletter
\newcommand\bigDiamond{\mathop{\mathpalette\bigDi@mond\relax}}
\newcommand\bigDi@mond[2]{%
  \vcenter{\hbox{\m@th
    \scalebox{\ifx#1\displaystyle 2\else1.2\fi}{$#1\Diamond$}%
  }}%
}
\makeatother

\usepackage{listings}
\title{A Proof System with Causal Labels (Part II): checking Counterfactual Fairness}

\author{Leonardo Ceragioli and Giuseppe Primiero\footnote{LUCI Lab, Department of Philosophy, Università degli Studi di Milano}} 

\date{}

\begin{document}

\maketitle

\begin{abstract}
In this article we propose an extension to the typed natural deduction calculus \textbf{TNDPQ} to model verification of counterfactual fairness in probabilistic classifiers. This is obtained formulating specific structural conditions for causal labels and checking that evaluation is robust under their variation.
\end{abstract}


\section{Introduction}

The calculus \textbf{TPTND} (\textit{Trustworthy Probabilistic Typed Natural Deduction} \cite{10.1093/logcom/exaf003,KUBYSHKINA2024109212}) is designed to evaluate \textit{post-hoc} the trustworthiness of the behavior of opaque systems. The system is implemented for verification of dataframes in the tool BRIO \cite{DBLP:conf/beware/CoragliaDGGPPQ23,coraglia2024evaluatingaifairnesscredit}.
In \cite{CeragioliPrimiero2025}, we introduced \textbf{TNDPQ} (\textit{Typed Natural Deduction for Probabilistic Queries}), a variation of the previous system in which a probabilistic output is associated to a target variable when a Data Point -- consisting of a list of values attributions for a set of variables -- is provided.
In this paper, we extend this system with tools to verify counterfactual fairness of classifiers.

We start with a formal definition of classifiers.
Let $\mathscr{A}$ be a set of protected variables $ a_{1} , \ldots , a_{n}$, $\mathscr{X}$ be a (disjoint) set of non-protected variables $ x_{1} , \ldots , x_{n}$ and $t$ be a target variable.
Moreover, let $\mathscr{V}_{a_{i}}$ be a set of values $\alpha ^{i_{1}}, \alpha ^{i_{2}}, \ldots , \alpha ^{i_{j}}$ that $a_{i}$ can receive, $\mathscr{V}_{A}$ the set of all $\mathscr{V}_{a_{i}}$, $\mathscr{V}_{x_{i}}$ be a set of values $\beta ^{i_{1}}, \beta ^{i_{2}}, \ldots , \beta ^{i_{j}}$ that $x_{i}$ can receive, $\mathscr{V}_{X}$ the set of all $\mathscr{V}_{x_{i}}$, and $\mathscr{V}_{t}$ the set $\delta ^{1}, \delta ^{2}, \ldots , \delta ^{j}$ of values that $t$ can receive.
Let us use $ v_{1} , \ldots , v_{n}$ to denote elements of $\mathscr{A} \cup \mathscr{X}$ (that is, variables regardless of their protected or unprotected status), and $\gamma ^{i_{1}}, \ldots , \gamma ^{i_{j}}$ to denote the values that $v_{i}$ can receive.
We use $a_{i}:\alpha ^{i_{j}}$ (respectively $x_{i}:\beta ^{i_{j}}$) to express the \textit{judgment} that variable $a_{i}$ receives value $\alpha ^{i_{j}}$ (respectively, variable $x_{i}$ receives value $\beta ^{i_{j}}$), and $t:\delta ^{1}_{p_{1}}, \ldots ,\delta ^{j}_{p_{j}} $ to express the \textit{probabilistic judgment} that $\delta ^{1}, \ldots , \delta ^{j}$ are all the possible values that variable $t$ can receive and that, for $1 \leq k \leq j$, it receives value $\delta ^{k}$ with probability $p_{k}$.\footnote{
\label{note:1}
We assume that the values for $t$ (also for the elements of $\mathscr{A}$ and $\mathscr{X}$ but this is irrelevant here) are all mutually exclusive, and so $\sum _{k=1}^{j} p_{k}=1$.
Note that we make no assumption regarding whether $t\in \mathscr{A} \cup \mathscr{X} $ and so on whether $\mathscr{V}_{t} = \mathscr{V}_{a_{i}} $ or $ \mathscr{V}_{t} = \mathscr{V}_{x_{i}} $ for some $i$.
}
We use $\mathscr{J}^{\mathscr{A}}$ for the set of all the judgments about protected variables, $\mathscr{J}^{\mathscr{X}}$ for the set of all the judgments about non-protected variables, and $\mathscr{J}^{\mathscr{P}}$ for the set of all probabilistic judgments.
Moreover, we use $\sigma ^{\mathscr{A}}$ to express a set of judgments about protected variables such that each element of $\mathscr{A}$ receives at most one value, $\sigma ^{\mathscr{X}}$ to express a set of judgments about non-protected variables such that each element of $\mathscr{X}$ receives at most one value, $\Sigma ^{\mathscr{A}}$ to refer to the set of all $\sigma ^{\mathscr{A}}$, and $\Sigma ^{\mathscr{X}}$ to refer to the set of all $\sigma ^{\mathscr{X}}$.
$\sigma$ is used to express the union of a $\sigma ^{\mathscr{A}}$ and a $\sigma ^{\mathscr{X}}$, and $\Sigma$ is used to refer to the set of all $\sigma$.
More formally:
\[
\Sigma ^{\mathscr{A}} =_{def} \{\sigma ^{\mathscr{A}} \subseteq \mathscr{J}^{\mathscr{A}} \mid \forall i ( a_{i}: \alpha ^{i_{l}} \in \sigma ^{\mathscr{A}} \land a_{i}: \alpha ^{i_{m}} \in \sigma ^{\mathscr{A}} \rightarrow l=m) \}
\]
\[
\Sigma ^{\mathscr{X}} =_{def} \{\sigma ^{\mathscr{X}} \subseteq \mathscr{J}^{\mathscr{X}} \mid \forall i ( x_{i}: \beta ^{i_{l}} \in \sigma ^{\mathscr{X}} \land x_{i}: \beta ^{i_{m}} \in \sigma ^{\mathscr{X}} \rightarrow l=m) \}
\]
\[
\Sigma =_{def} \{\sigma ^{\mathscr{A}} \cup \sigma ^{\mathscr{X}} \mid \sigma ^{\mathscr{A}} \in \Sigma ^{\mathscr{A}} ~ \land ~ \sigma ^{\mathscr{X}} \in \Sigma ^{\mathscr{X}} \}
\]

\noindent
A classifier $\widehat{\mathscr{f}} \in \widehat{\mathscr{F}} $ is a function from $\Sigma$ to $\mathscr{J}^{\mathscr{P}}$, where each $\sigma \in \Sigma$ describes a Data Point, that is what we know about a subject, and the probabilistic judgment $t: \delta ^{1} _{p_{1}}, \ldots ,\delta ^{j} _{p_{j}} $ in $\mathscr{J}^{\mathscr{P}}$ represents the output of the classifier regarding the probability distribution of the possible values for the target variable $t$.

\textbf{TNDPQ} is a proof system working with sequents describing the result of queries for classifiers.
More precisely, each classifier $\widehat{\mathscr{f}}$ is characterized by a set of ground sequents of the form:\footnote{
Technically, we should add a subscript in the equation specifying the classifier we are focusing on.
However, this will not be needed here, since we will not compare outputs of different classifiers.}

\begin{equation}
\label{eq:sequents}
    \nc{\sigma}{t: \delta _{p}}
\end{equation}

\noindent
For readability reasons, the sequents focus on only one possible value for the target variable at a time.
In \cite{CeragioliPrimiero2025} we show how to extend \textbf{TNDPQ} with sequents working with logically complex sequents -- possibly non-atomic variables receiving possibly non-atomic values.
As an example, the following sequent expresses the probability that a non-white 27 years old man who is married or divorced and has a gross annual income of $65000 €$ receives a loan:
\[
\nc{Age: 27 ,\: Gen.: m,\: MS: married + divorced,\: Etn.: white^{\bot},\: GAI : 65K}{Loan : yes _{0.60}}
\]
\indent \textbf{TNDPQ} was initially designed to investigate the preservation of trustworthiness under the composition of logically simpler queries and then extended with causal labels to verify individual and intersectional fairness for a probabilistic classifier via structural properties, see \cite{ceragioli_primiero2025}.
In this paper, we focus only on the atomic fragment and provide a \textit{criterion} to verify counterfactual fairness for a probabilistic classifier.

\section{Counterfactual Fairness}

Counterfactual fairness requires that a subject would not have been treated differently had their protected attributes been different \cite{kusner2017a}.
Formally, it can be defined as follows:

\begin{definition}[Counterfactual Fairness (\textbf{CF})]
    An algorithm is counterfactually fair regarding a protected variable $a$ if, given a Data Point $\sigma$ describing an actual individual, the algorithm gives the same output to both $\sigma$ and to the Data Point $\sigma '$ describing how the individual would have been, had the protected variable $a$ received a different value.
\end{definition}

As an example, we could wonder whether the probability of receiving a loan in the previous example would have still been $60\%$, had the subject been a woman?
If we do not consider the connections between the features, this question just corresponds to whether or not the following sequent is derivable in the system:\footnote{
Note that this would make counterfactual and individual fairness indistinguishable from one another.}
\[
\nc{Age: 27 ,\: Gen.: f,\: MS: married + divorced,\: Etn.: white^{\bot},\: GAI : 65K}{Loan : yes _{0.60}}
\]

\noindent
However, when causal relations are taken into account, it is a trivial observation that gender influences other features (both directly and indirectly).
For example, both through objective physical differences and prejudices, gender influences job opportunities, and therefore the counterpart of a subject having a different gender would probably not have the same $GAI$.
This makes the individuation of the Data Point $\sigma '$ a lot more complicated.

Hence, assessing counterfactual situations \textbf{CF} requires some more precise formal tools.
The usual approaches are: possible worlds semantics \cite{Lewis1973-LEWC-2} and causal models \cite{pearl2017}.
In the following we choose the second method.

\section{Causal Relations}

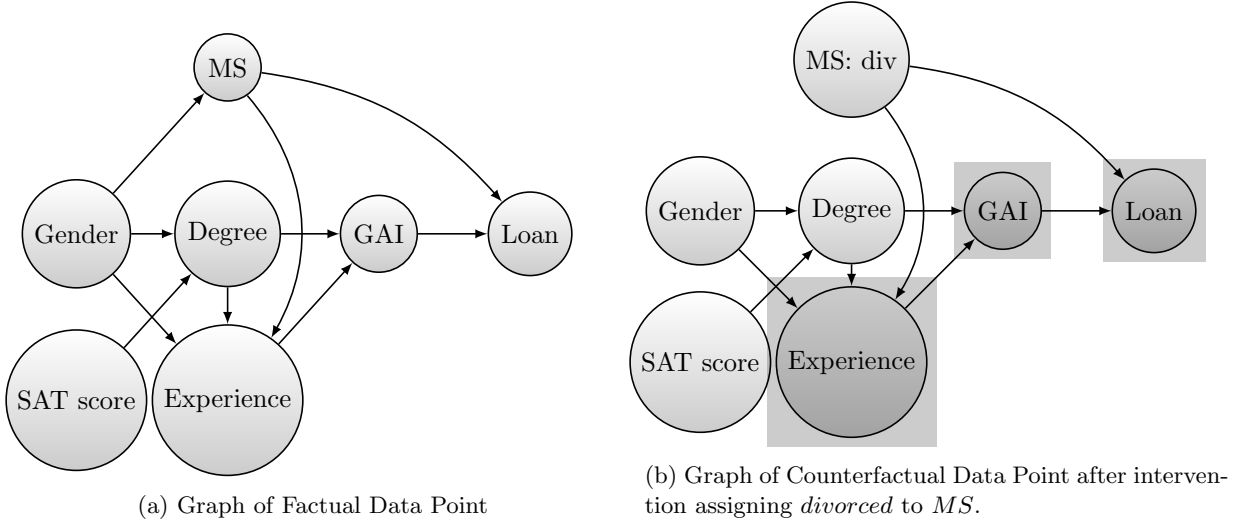
\begin{figure}
\begin{subfigure}[b]{0.49\textwidth}
\captionsetup{width=0.95\textwidth}
\begin{tikzpicture}[-latex ,auto ,node distance =2.2 cm and 2 cm ,on grid ,
semithick ,
state/.style ={ circle ,top color =white , bottom color = black!20 ,
draw, text=black , minimum width =0.5 cm}]

\node[state] (1) {Gender};
\node[state] (2) [above right=of 1] {MS};
\node[state] (3) [right=of 1] {Degree};
\node[state] (4) [below right=of 1] {Experience};
\node[state] (5) [right=of 3] {GAI};
\node[state] (6) [right=of 5] {Loan};
\node[state] (7) [below=of 1] {SAT score};

\path (1) edge (2);
\path (1) edge (3);
\path (1) edge (4);
\path (3) edge (4);
\path (3) edge (5);
\path (4) edge (5);
\path (2) edge [bend left =20] (6);
\path (5) edge (6);
\path (7) edge (3);
\path (2) edge [bend left =35] (4);

\end{tikzpicture}
\caption{Graph of Factual Data Point}
\end{subfigure}
\begin{subfigure}[b]{0.49\textwidth}
\captionsetup{width=0.95\textwidth}
\begin{tikzpicture}[-latex ,auto ,node distance =2 cm and 2 cm ,on grid ,
semithick ,
state/.style ={ circle ,top color =white , bottom color = black!20 ,
draw, text=black , minimum width =0.5 cm}]

\node[state] (1) {Gender};
\node[state] (2) [above right=of 1] {MS: div};
\node[state] (3) [right=of 1] {Degree};
\node[state] (4) [below right=of 1] {Experience};
\node[state] (5) [right=of 3] {GAI};
\node[state] (6) [right=of 5] {Loan};
\node[state] (7) [below=of 1] {SAT score};

\node[fit=(4), fill=black, opacity=0.2] {};
\node[fit=(5), fill=black, opacity=0.2] {};
\node[fit=(6), fill=black, opacity=0.2] {};

\path (1) edge (3);
\path (1) edge (4);
\path (3) edge (4);
\path (3) edge (5);
\path (4) edge (5);
\path (2) edge [bend left =20] (6);
\path (5) edge (6);
\path (7) edge (3);
\path (2) edge [bend left =35] (4);

\end{tikzpicture}
\caption{Graph of Counterfactual Data Point after intervention assigning $divorced$ to $MS$.}
\end{subfigure}

\caption{Graphs corresponding to factual and counterfactual Data Points.
The squares in figure (b) surround variables that depend on the variable we intervene on, and which for this reason cannot be used to decide the target.
}
\label{fig:intervention}
\end{figure}

We first consider how counterfactual situations are dealt with in causal models, and then internalize both causal relations and the characterization of counterfactuals in our calculus \textbf{TNDPQ}.
For the purposes of this work, we define causal graphs as follows:

\begin{definition}[Causal Graph]
    A causal graph is an acyclic directed graph with nodes representing events (variables receiving values) and edges representing immediate causal relations.
\end{definition}

By closing edges under transitivity, we obtain the notion of mediate cause.
For purely formal reasons, we close the notion of cause under reflexivity as well.
The usual extension of deterministic causal graphs with functions to compute the value of a node on the basis of those of all the immediate parent nodes is here expressed by sequents like in equation~\ref{eq:sequents}.\footnote{
However, note that in causal models the function does not assign value in a probabilistic way and probabilities come in play only at a later stage.
On the contrary, our equations are probabilistic in the strictest sense. 
}

As already shown, to capture counterfactuals one cannot just change the protected attribute and leave all other variables fixed: the properties of the subject that do not depend on the protected variable need to be identified and kept fixed.
For this, the usual approach (followed, for example, in \cite{pearl2017,kusner2017a}) is to rely on the distinction between exogenous and endogenous variables: exogenous variables represent attributes that have no direct cause in the graph, while endogenous ones represent their consequences.
By keeping fixed the values of exogenous variables (possibly with the exclusion of the protected variable), we make sure that the causal graph represents the counterfactual situation.
In fact, exogenous variables cannot be consequences of the protected variable.
Moreover, under the assumptions of causal models, it is possible to calculate the values of the endogenous variables from those of the exogenous ones.

According to this usual approach, the protected variable can be either exogenous (such as gender) or endogenous (such as marital status, which, for example, depends on gender: there are more widows than widowers).
When the protected variable is endogenous, we erase all the edges that enter it, since we want to assign its value \textit{ad arbitrium}.

In summary, to capture counterfactual situations, the usual approach prescribes to intervene on the graph as follows \cite{pearl2017,kusner2017a}: 

\begin{enumerate}
    \item \textit{impose} some value to the protected variable; \item \textit{keep} all the (other) exogenous variables fixed; \item \textit{erase} all the edges that enter in the protected variable; \item \textit{calculate} the values of the endogenous variables (particularly of the target), using those of the exogenous variables.

\end{enumerate}
The third point is relevant only when the protected variable is not exogenous; otherwise, it is vacuously satisfied.
Now, to check \textbf{CF} we just need to control whether the resulting value of the target variable is the same.

We slightly modify this approach to apply it to ML classifiers.
The assignment of a value to the target variable depends on the selection of a set of exogenous variables, so we have to be sure to work with an adequate set of such variables.
Moreover, while in causal models we can require to have a sufficiently vast set of exogenous variables, we cannot ask the same regarding the set of entries of a classifier, which we cannot change.
For this reason, instead of relying only on the exogenous variables, we will use all and only the variables that are not (direct or indirect) effects of the protected one, and ignore all the others.
This will allow to use all the factual information that remains valid in the counterfactual situation in order to derive the counterfactual classification.
An example of the graphs corresponding to a factual and a counterfactual Data Point is shown in figure~\ref{fig:intervention}.

\section{Adding Counterfactuals to TNDPQ}

To express the fact that a Data Point is the counterfactual of another, we need to internalize both causal relations and interventions in our calculus \textbf{TNDPQ}.
For this purpose, we use the methodology of labeled calculi \cite{Vigano2000,NegriVonPlato2011}.
First, we extend the language with the following relational predicates for variables and expressions for interventions on Data Points, as already introduced in \cite{ceragioli_primiero2025}:

\begin{description}
\item[Immediate Causal Relations] $ v_{i} \tright v_{j} =_{def}$ $v_{i}$ is an immediate cause of $v_{j}$.

\item[Mediate Causal Relations] $ v_{i} \btright ^{M} v_{j} =_{def}$ $v_{i}$ is a mediate cause of $v_{j}$, with intermediate causes $M$.

\item[Intervention on Data Point] $[\tright _{\widehat{\mathscr{f}}}~, \sigma] I(a:\alpha) =_{def}$ an intervention assigning the value $\alpha$ to variable $a$ is operated on the Data Point $\sigma $ and its associated graph $\tright _{\widehat{\mathscr{f}}}$.\footnote{
Notice that we focus on interventions on protected variables, although interventions on unprotected variables could also be studied.
}
\end{description}

\noindent
Then, we reformulate \textbf{TNDPQ} sequents by extending their left-hand side:

\begin{equation}
\label{eq:counterfactual}
    \nc{\tright _{\widehat{\mathscr{f}}}~, \sigma}{t: \delta _{p}} \qquad \nc{ [\tright _{\widehat{\mathscr{f}}}~,\sigma ] I(a: \alpha) }{ t: \delta _{q}}
\end{equation}

Let us use $Var_{\sigma}$ to indicate the set of variables that occur in $\sigma$. 
We use $\tright _{\widehat{\mathscr{f}}}~$ to indicate all the immediate causal relations among features in the classifier.
$\btright _{\widehat{\mathscr{f}}}~$ denotes all the mediate causal relations in the resulting graph and is derivable as the closure of $\tright _{\widehat{\mathscr{f}}}~$ under reflexivity and transitivity.
We will use $\tright _{\widehat{\mathscr{f}}}'~$ and $\btright _{\widehat{\mathscr{f}}}'~$ respectively for different sets of direct and indirect causal relations.
Hence, the first sequent of equation~\ref{eq:counterfactual} is a sequent that gives an output for the target $t$ in the actual situation (that is, $\sigma$), also specifying the immediate causal relations that hold between the features of the classifier ($\tright _{\widehat{\mathscr{f}}}$), while the second is a hypersequent that gives an output for the target $t$ in the counterfactual situation resulting from $\tright _{\widehat{\mathscr{f}}}~,\sigma$ by the intervention that assigns $\alpha$ to $a$.
We call $a$ the variable of intervention.
Although $\tright _{\widehat{\mathscr{f}}}~$ is not actually used by the classifier to evaluate $t$, it is relevant in the calculus to check whether a Data Point is the counterfactual of another.

\begin{example}[Factual and Counterfactual sequents]
The following sequents express, respectively, that the probability of receiving a loan for a 27 years old person with a gross annual income of $40.000 €$ is 60\%, and that it would have been 50\% had them been 35 years old:
\[
\nc{Age \tright MS ,Age \tright GAI , Age \tright Loan, GAI \tright Loan, Age: 27 ,\: GAI: 40K }{Loan : yes _{0.60}}
\]
\[
\Scale[0.9]{\nc{[Age \tright MS ,Age \tright GAI , Age \tright Loan, GAI \tright Loan, Age: 27 ,\: GAI: 40K]I(Age: 35) }{Loan : yes _{0.50}}}
\]    
\end{example}

\begin{table*}
\centering
  \caption{Rules for the counterfactual, with the following conditions: (*) $ v_{k} \neq a $; (**) $v_{i}$ s.t. $v_{i} : \gamma ^{i} \in \sigma '  $ and for no set of points $M$, $ a \btright ^{M} v_{i} \in \: \btright _{\widehat{\mathscr{f}}}'~$.}
  \label{tab:counterfactual}
\begin{tabular}{c c}
\AxiomC{{\small $ \nc{ \tright _{\widehat{\mathscr{f}}}~,\sigma}{t: \delta _{p}} $}}
	\RightLabel{{\tiny C-Weakening}}
	\UnaryInfC{{\small $ \nc{ [\tright _{\widehat{\mathscr{f}}} '~,\sigma '] I(a: \alpha), \tright _{\widehat{\mathscr{f}}}~,\sigma }{ t: \delta _{p}} $}}
\DisplayProof
&
\AxiomC{{\small $ \nc{ [\tright _{\widehat{\mathscr{f}}} '~,\sigma '] I(a: \alpha), \tright _{\widehat{\mathscr{f}}}~, \sigma , a: \alpha}{ t: \delta _{p}} $}}
	\RightLabel{{\tiny $I$-Cut}}
	\UnaryInfC{{\small $ \nc{ [\tright _{\widehat{\mathscr{f}}} '~,\sigma '] I(a: \alpha), \tright _{\widehat{\mathscr{f}}}~,\sigma }{ t: \delta _{p}} $}}
\DisplayProof
\\[0.7cm]
\AxiomC{{\small $ \nc{ [\tright _{\widehat{\mathscr{f}}} '~,\sigma '] I(a: \alpha), \tright _{\widehat{\mathscr{f}}}~, v_{i}\tright v_{k}~, \sigma }{ t: \delta _{p}} $}}
	\RightLabel{{\tiny $\tright$-Cut}$^{*}$}
	\UnaryInfC{{\small $ \nc{ [\tright _{\widehat{\mathscr{f}}} '~,\sigma '] I(a: \alpha), \tright _{\widehat{\mathscr{f}}}~,\sigma }{ t: \delta _{p}} $}}
\DisplayProof
&
\AxiomC{{\small $ \nc{ [\tright _{\widehat{\mathscr{f}}} '~,\sigma '] I(a: \alpha), \tright _{\widehat{\mathscr{f}}}~,\sigma , v_{i}:\gamma ^{i} }{ t: \delta _{p}} $}}
	\RightLabel{{\tiny $v$-Cut}$^{**}$}
	\UnaryInfC{{\small $ \nc{ [\tright _{\widehat{\mathscr{f}}} '~,\sigma '] I(a: \alpha), \tright _{\widehat{\mathscr{f}}}~,\sigma }{ t: \delta } _{p}$}}
\DisplayProof
\\
\end{tabular}
\end{table*}

While sequents describing actual decisions of a classifier, such as the one on the left of equation~\ref{eq:counterfactual}, are assumptions of \textbf{TNDPQ}, those describing counterfactual decisions, such as the one on the right of the equation, are derivable.
To derive them, we start with a plausible sequent for the counterfactual, which can be obtained using only the features that do not depend on the variable of intervention to run the classifier:
\[
\nc{ \tright _{\widehat{\mathscr{f}}} '~,\sigma '}{ t: \delta _{q}}
\]

\noindent
Let us call this the counterfactual candidate.
Then, we apply the rule C-Weakening in table~\ref{tab:counterfactual}, adding $[\tright _{\widehat{\mathscr{f}}}~,\sigma]I(a:\alpha)$.
The resulting hypersequent $\nc{[\tright _{\widehat{\mathscr{f}}}~,\sigma] I(a: \alpha) \tright _{\widehat{\mathscr{f}}} '~,\sigma '}{ t: \delta _{q}}$ can be interpreted as saying that the classifier assigns probability $q$ to the value assignment $t: \delta$ for the Data Point $\sigma '$, which we regard as a counterfactual candidate for the Data Point $\sigma$ after intervention assigning to $a$ the value $\alpha$.
The formulas $\tright _{\widehat{\mathscr{f}}}~$ and $\tright _{\widehat{\mathscr{f}}}'~$ represent, respectively, the graph describing the causal relations between the variables of the classifier and the same graph after the intervention.

If $\tright _{\widehat{\mathscr{f}}} '~,\sigma '$ is really the counterfactual of $\tright _{\widehat{\mathscr{f}}}  ,\sigma $ after intervention $I(a:\alpha)$, by applying the rules of Cut in table~\ref{tab:counterfactual} we end with a sequent of the form:
\[
\nc{ [\tright _{\widehat{\mathscr{f}}}~,\sigma] I(a: \alpha) }{ t: \delta _{q}}
\]
\noindent
More precisely, the rule $I$-Cut erases $a: \alpha$ from the premise, that is, the value assignment to the protected variable imposed by the intervention.
Hence, if the counterfactual candidate contains $a: \alpha$, it can be erased.
The rule $\tright$-Cut erases $v_{i}\tright v_{k}$, under the condition that $ k \neq j $, enabling the erasure of all direct causal relations that do not enter the protected variable.
The rule $v$-Cut erases $v_{i}:\gamma ^{i} $, that is, the assignment of value given to $v_{i}$ by the original Data Point, under the condition that $v_{i}$ is not a consequence of $ v_{j} $.
This is established by checking $\btright _{\widehat{\mathscr{f}}}$, that is, the set of mediate causal relations resulting from the graph before intervention.
Hence, all the assignments of values to the variables that are not consequences of the protected variable in the original graph can be erased.

The label $Cut$ of these rules refers to the fact that they can be seen as contractions of Cut applications of the following kind:\footnote{
See note~\ref{note:1} for the occurrence of $a$ as target variable.
}

\begin{center}
    \AxiomC{$\nc{ [\tright _{\widehat{\mathscr{f}}} '~,\sigma '] I(a: \alpha)}{a: \alpha _{1}}$}
    \AxiomC{$\nc{\tright _{\widehat{\mathscr{f}}}~, \sigma , a: \alpha}{ t: \delta _{p}}$}
	\RightLabel{{\small Cut}}
	\BinaryInfC{$\nc{ [\tright _{\widehat{\mathscr{f}}} '~,\sigma '] I(a: \alpha), \tright _{\widehat{\mathscr{f}}}~,\sigma }{t: \delta _{p}}$}
\DisplayProof
\end{center}

\noindent
Where the first premise says that in the counterfactual Data Point obtained from $\tright _{\widehat{\mathscr{f}}} '~,\sigma '$ by the intervention $I(a: \alpha )$, variable $a$ receives the value $\alpha$ with probability $1$ (that is, with certainty), and the second premise gives the probability of $t: \delta$ in the counterfactual candidate.
$\tright$-Cut and $v$-Cut are contractions of similar Cut applications.

If by applying all these rules we can erase all the formulas in the counterfactual candidate, then this is really the Data Point corresponding to the counterfactual of the factual Data Point.
Now, all we have to do to check \textbf{CF} is to compare the probabilities $p$ and $q$.
This can be done either by requiring their identity or by requiring a threshold on their difference.

\begin{example}[Evaluation of \textbf{CF} for a classifier]
Let us use $\tright _{\widehat{\mathscr{f}}}$ for the causal relations resulting from graph (a) of figure~\ref{fig:intervention} and $\tright _{\widehat{\mathscr{f}}}'$ for the causal relations resulting from graph (b) of the same figure.
Moreover, let us assume that the following sequent describe decisions of a classifier:
\begin{equation}
\label{eq:ex1}
    \Scale[0.9]{\nc{\tright _{\widehat{\mathscr{f}}}~,G.: m,\: MS: mar ,\: SAT: 1100,\: GAI : 65K,\: Deg.: PhD ,\: Exp: 5 y}{Loan : yes _{0.60}}}
\end{equation}
\begin{equation}
\label{eq:ex2}
 \Scale[0.9]{\nc{\tright _{\widehat{\mathscr{f}}} '~,G.: m,\: MS: div,\: SAT: 1100,\: Deg.: PhD }{Loan : yes _{0.60}}}
\end{equation}
We can show that sequent~\ref{eq:ex2} is the counterfactual of \ref{eq:ex1} after intervention $I(MS:div)$, since it entails the sequent:
\begin{equation}
\label{eq:ex3}
 \Scale[0.85]{\nc{[\tright _{\widehat{\mathscr{f}}}~,G.: m,~ MS: mar ,~ SAT: 1100,~ GAI : 65K,~ Deg.: PhD ,~ Exp: 5 y]I(MS:div)}{Loan : yes _{0.60}}}
\end{equation}
The derivation can be constructed as follows, using $[\ldots]$ for $[\tright _{\widehat{\mathscr{f}}}~,G.: m,~ MS: mar ,~ SAT: 1100,~ GAI : 65K,~ Deg.: PhD ,~ Exp: 5 y]$ and relying only on the rules in table~\ref{tab:counterfactual}:
\begin{center}
    \AxiomC{{\footnotesize $\nc{\tright _{\widehat{\mathscr{f}}} '~,G.: m,\: MS: div,\: SAT: 1100,\: Deg.: PhD }{Ln : y _{0.60}}$}}
    \RightLabel{{\tiny C-W}}
	\UnaryInfC{{\footnotesize $\nc{[\ldots]I(MS:div) ,\tright _{\widehat{\mathscr{f}}} '~,G.: m,\: MS: div,\: SAT: 1100,\: Deg.: PhD }{Ln : y _{0.60}}$}}
    \RightLabel{{\tiny $I$-Cut}}
	\UnaryInfC{{\footnotesize $\nc{[\ldots]I(MS:div) ,\tright _{\widehat{\mathscr{f}}} '~,G.: m,\: SAT: 1100,\: Deg.: PhD }{Ln : y _{0.60}}$}}
    \RightLabel{{\tiny $\tright$-Cut}}
    \doubleLine
	\UnaryInfC{{\footnotesize $\nc{[\ldots]I(MS:div) ~,G.: m,\: SAT: 1100,\: Deg.: PhD }{Ln : y _{0.60}}$}}
    \RightLabel{{\tiny $v$-Cut}}
    \doubleLine
	\UnaryInfC{{\footnotesize $\nc{[\tright _{\widehat{\mathscr{f}}}~,G.: m,~ MS: mar ,~ SAT: 1100,~ GAI : 65K,~ Deg.: PhD ,~ Exp: 5 y]I(MS:div)}{Ln : y _{0.60}}$}}
\DisplayProof
\end{center}

\noindent
Moreover, since the probability that the variable $Ln$ receives value $y$ is the same in both sequents, this application of the classifier satisfies \textbf{CF}.
\end{example}

\section{Conclusion}

This work focuses on formal tools to check counterfactual fairness of probabilistic classifiers.
We have seen how causal models allow to formalize counterfactuals, and argued that a modified version of their approach is suitable to test fairness for classifiers.
We have proposed a typed natural deduction calculus \textbf{TNDPQ} extended with labels representing causal relations and expressions for interventions to internalize this approach.

\section*{Acknowledgments}

This research was supported by the Ministero dell’Università e della Ricerca (MUR) through PRIN 2022 Project SMARTEST – Simulation of Probabilistic Systems for the Age of the Digital Twin (20223E8Y4X), and through the Project “Departments of Excellence 2023-2027” awarded to the Department of Philosophy “Piero Martinetti” of the University of Milan.

\section*{Declaration on Generative AI}

 During the preparation of this work, the authors used Grammarly in order to: Grammar and spelling check. After using these tool, the authors reviewed and edited the content as needed and take full responsibility for the publication’s content.

\bibliography{biblio}


\end{document}